\documentclass[aps,twocolumn,superscriptaddress]{revtex4-1}
\usepackage{graphicx,hyperref,epstopdf,amsfonts,amsmath}
\usepackage{amssymb}
\usepackage{color}

\begin{document}

\title{Molecular search with conformational change: One-dimensional discrete-state stochastic model}
\author{Jaeoh Shin}
\affiliation{Department of Chemistry, Rice University, Houston, Texas, 77005, USA}
\author{Anatoly B. Kolomeisky}
\affiliation{Department of Chemistry, Rice University, Houston, Texas, 77005, USA}
\affiliation{Department of Chemical and Biomolecular Engineering, Rice University, Houston, Texas, 77005, USA}
\affiliation{Center for Theoretical Biological Physics, Rice University, Houston, Texas, 77005, USA}
\date{July 5, 2018}

\begin{abstract}

Molecular search phenomena are observed in  a variety of chemical and biological systems. During the search the participating particles frequently move in complex inhomogeneous environments with random transitions between different dynamic modes. To understand the mechanisms of molecular search with alternating dynamics, we investigate the search dynamics with stochastic transitions between two conformations in a one-dimensional discrete-state stochastic model. It is explicitly analyzed using the first-passage time probabilities method to obtain a full dynamic description of the search process. A general dynamic phase diagram is developed. It is found that there are several dynamic regimes in the molecular search with conformational transitions, and they are determined by the relative values of the relevant length scales in the system. Theoretical predictions are fully supported by Monte Carlo computer simulations.

\end{abstract}
	
\maketitle
\section{Introduction}
\label{intro}

Many chemical and biological processes involve molecular search that frequently takes place in complex inhomogeneous media, leading to random changes in the dynamic properties of the participating particles \cite{benichou11review,bressloff13}. For instance, in the heterogeneous chemical catalysis the reacting molecules alternate between the fluid phase and the solid surface of the catalyst before the reaction is completed \cite{ross_book}. Similar alternating dynamics is observed in gene activation by proteins, where the transcription factors associate to  specific sequences on DNA \cite{gorman08}. Experimental studies showed that this process is a combination three-dimensional (3D) diffusion in the bulk solution and one-dimensional  (1D) sliding along the DNA chain \cite{gorman08}. Such molecular search processes with alternating dynamics are known as {\it intermittent searches}, and in recent years they have been intensively studied both experimentally and theoretically \cite{adam68,riggs70,berg76,berg81,berg89,mirny04,kolomeisky08,mirny09,kolomeisky11,bauer12,metzler17,benichou11review,bressloff13,godec17}.

The most intriguing feature of the processes with the intermittent search is the random switching of the molecules between different dynamic states. This observation raises several fundamental questions on the mechanisms of such complex processes. Why natural processes frequently exhibit such complex dynamics? How specifically these stochastic transitions affect the dynamics? Is the overall dynamics always accelerated by these alternating dynamics? These questions have been discussed before, but the overall molecular picture for the mechanisms of the processes with the intermittent search remains not fully understood \cite{benichou11review,kolomeisky11,sheinman12,bressloff13}.

One of the simplest intermittent search systems is the 1D case where the reacting molecule searches for the target while moving along a line and alternating between several conformations with different dynamic properties. This model is relevant for the understanding of how the transcription factors already bound to DNA can locate the specific target sequences. It was shown experimentally that the protein-DNA complexes have different conformations, leading to the variable strength of the protein-DNA interactions during the search process \cite{lac04, p53exp,iwahara13}.
When the protein molecule weakly interacts with the DNA chain, it can diffuse rapidly, and this is known as a {\it search} conformation. But the target can be found only in a so-called {\it recognition} conformation when the protein interacts stronger with DNA while diffusing slower. Apparently, alternating between two conformations might help proteins to find the target faster at some conditions \cite{kochugaeva17}.

The molecular search with conformational transitions in 1D systems was theoretically considered before \cite{benichou07, holcman09,holcman10,holcman11,holcman15,ivanov14,kochugaeva16}. In Ref. \cite{holcman09,holcman10,holcman11,holcman15}, the authors presented a continuum model, where the searcher diffuses with alternating diffusivity in the continuum space. However, the analysis was done only for the search times at one specific location of the target. In addition, in the limit of very small diffusivity in the recognition conformational state, the search times become infinitely large, which is physically unreasonable.
This shows that the application of this approach for real biological processes is problematic at these conditions. It has been argued before that a more general discrete-state description is required to properly analyze biochemical processes with search \cite{veksler13}. In Ref. \cite{benichou07}, the search dynamics in both continuous and discrete space was analyzed. It was assumed that the molecule can switch between the two regimes: a diffusion conformation from which the target can be found and a so-called teleportation conformation where the searcher can jump to any location in the system but the target cannot be found.  However, the motion in the fast teleportation mode was not considered explicitly, and this also limited the applicability of this model. In Ref. \cite{ivanov14}, the authors considered an infinite lattice of discrete states which are visited by the searching molecules. But the explicit analysis was done only in the limiting cases with the assumption of conformational equilibrium  which restricts the applicability of this approach to real biological systems.
Our previous studies of this process investigated 1D conformational changes coupled to the bulk diffusion and only limiting cases were analyzed \cite{kochugaeva16}.

In this paper, we present a general theoretical approach to describe the molecular search with conformational transitions in one dimension. The searcher molecule stochastically transitions between two dynamic modes with different hopping rates on the lattice, and it can recognize the target only in one of the conformational states. Using a method of first-passage probabilities, a full dynamic description of the system is obtained, and we concentrate on explicit calculations of the mean search times. By exploring all parameter ranges, several dynamic regimes are identified and described. It is argued that these different regimes are specified by the dominating length scales in the system. Our theoretical analysis is also supported by Monte Carlo computer simulations.

The paper is organized as follows. We introduce the model in Sec. \ref{model} and the main results are presented in Sec. \ref{results}. In Sec. \ref{approx} we identify two limiting cases for which physically transparent solutions are obtained. Finally, we summarize and conclude in Sec. \ref{conclusion}. The details of the calculations are presented in the Appendix.

\section{Theoretical Model}
\label{model}

\begin{figure}
\includegraphics[width=7.5cm]{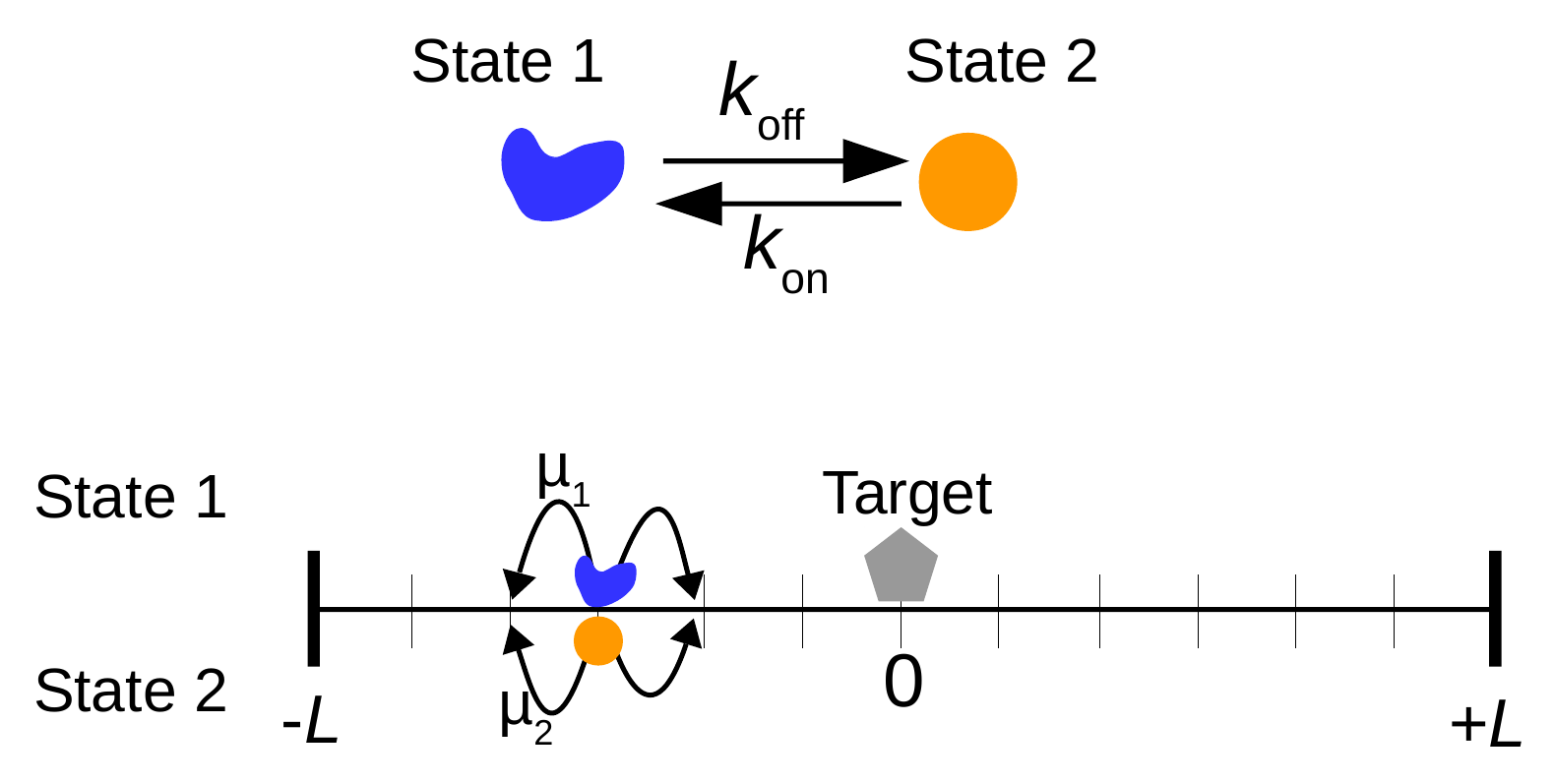}
\caption{A schematic view of the molecular search for a target with conformational change. (Top) The molecule switches between two states and it can recognize target only when it is in state 1. The transition rates between two states are $k_{\text{on}}$ and $k_{\text{off}}$. (Bottom) The hopping rates on the lattice are $\mu_1$ and $\mu_2$ depending on the state of molecule. }
\label{fig-1}
\end{figure}

We consider the search of a single molecule for a target site in the one-dimensional lattice of discrete sites as shown in Fig. \ref{fig-1}. While the immobile target is located at a single site of the lattice,
the searcher molecule diffuses along the lattice. During the process, the molecule can randomly switch between two conformational states: the recognition mode and the search mode. In the recognition mode (labeled as a state 1), the molecule can recognize the target site, and it moves with the hopping rate $\mu_1$ with equal probability in both directions. In the search mode (labeled as a state 2), it cannot recognize the target, but its diffusional hopping rate in this state is $\mu_2$, which is generally different from the rate $\mu_1$. The stochastic switchings between two conformations are assumed to be Poissonian with the rate $k_{\text{on}}$ (from the search mode to the recognition mode) and $k_{\text{off}}$ (from recognition mode to search mode), see Fig. {\ref{fig-1}. The number of lattice sites is $2L+1$ which is labeled as $n=-L$, $-L+1$, ..., 0, ... $L-1$, $L$. We consider reflecting boundary conditions on both sides of the lattice, and for convenience the target is located at the center ($n=0$), although our analysis can be extended for any location along the lattice.

To describe the search dynamics in this system, we employ the method of first-passage probabilities that was successful in analyzing various problems related to protein search for targets on DNA \cite{veksler13,shvets15,kochugaeva16,kochugaeva17,shin18}. We start by defining the first-passage time probability density function $F_{i}(n,t)$ to reach the target at time $t$ given that the molecule was at site $n$ as in the $i$-th state ($i$= 1 or 2) at time $t=0$. The temporal evolution of the first-passage probability functions is governed by the backward master equations \cite{redner},
\begin{widetext}
\begin{equation}
	\frac{\partial F_1(n,t)}{\partial t}=\mu_1[F_1(n-1,t)+F_1(n+1,t)]+k_{\text{off}}F_2(n,t) - (2\mu_1 + k_{\text{off}})F_{1}(n,t), 
\end{equation}\label{eq1}
\begin{equation}
	\frac{\partial F_2(n,t)}{\partial t}=\mu_2[F_2(n-1,t)+F_2(n+1,t)]+k_{\text{on}}F_1(n,t)-(2\mu_{2}+k_{\text{on}})F_2(n,t),
\end{equation}\label{eq2}
\end{widetext}
for $-L < n < L$. The physical meaning of these equations is that all trajectories starting on the site $n$ and reaching the target can be divided into three groups: going first to the site $n-1$, going first to the site $n+1$ or switching first to another conformation in the site $n$. At the boundaries ($n=\pm L$), the equations are slightly different to reflect the geometry of these locations (see the Appendix). The initial condition is $F_1(n=0,t)=\delta(t)$, which means that if the initial position of the molecule is at the target site in the conformational state 1, it finds the target instantaneously.

To solve these equations we apply the Laplace transform, $\widetilde{F_i(n,s)}=\int_{0}^{\infty}F_i(n,t)\exp(-st)dt$, where $s$ is the Laplace variable. Then the above master equations can be rewritten as
\begin{widetext}
\begin{equation}
(s+2\mu_1+k_{\text{off}})\widetilde{F_1(n,s)}=\mu_1[\widetilde{F_1(n-1,s)}+\widetilde{F_1(n+1,s)}]+k_{\text{off}}\widetilde{F_2(n,s)},
\label{eq3}
\end{equation}
\begin{equation}
(s+2\mu_2+k_{\text{on}})\widetilde{F_2(n,s)}=\mu_2[\widetilde{F_2(n-1,s)}+\widetilde{F_2(n+1,s)}]+k_{\text{on}}\widetilde{F_1(n,s)}.
\label{eq4}
\end{equation}
\end{widetext}
These coupled second-order difference equations can be solved by transforming them into a single fourth-order difference equation. The details of the calculations are presented in the Appendix. 
Once we obtain $\widetilde{F_i(n,s)}$, it is straightforward to compute all dynamic properties in the system. Let us concentrate on the mean search times, or mean-first passage times, which can be obtained from
\begin{equation}
	T_i(n)=\int_{0}^{\infty} t F_i(n,t)dt=-\frac{\partial\widetilde{F_i(n,s)}}{\partial s}\vert_{s=0}.
\end{equation}

\section{Results and Discussion}
\label{results}

Our theoretical method allows us to calculate the mean search times $T_{i}(n)$ for arbitrary kinetic parameters, for any location of the target and for any lattice size $L$. To be specific, let us put the target in the middle of the lattice ($n=0$), and assume that at $t=0$ the molecule with equal probability can start from any site in the search conformation ($i=2$). We will evaluate the position average search time,
\begin{equation}
T= \langle T_{2}(n)\rangle =\frac{1}{2L+1}\sum_{n=-L}^{L}T_{2}(n),
\end{equation}
for various sets of kinetic parameters. It is important to note that there are four length scales in the system that are governing the search dynamics. The first one is the size of the lattice $L$, the next one is the size of the target, which takes the single lattice site,
and the last two are scanning lengths $\lambda_1$ and $\lambda_2$, which are defined as $\lambda_1\equiv\sqrt{\mu_1/k_{\text{off}}}$ and $\lambda_2\equiv\sqrt{\mu_2/k_{\text{on}}}$. The physical meaning of these scanning lengths is the average distance that the molecule moves along the lattice in one conformational state before switching to the other state.

\begin{figure}
\includegraphics[width=7.5cm]{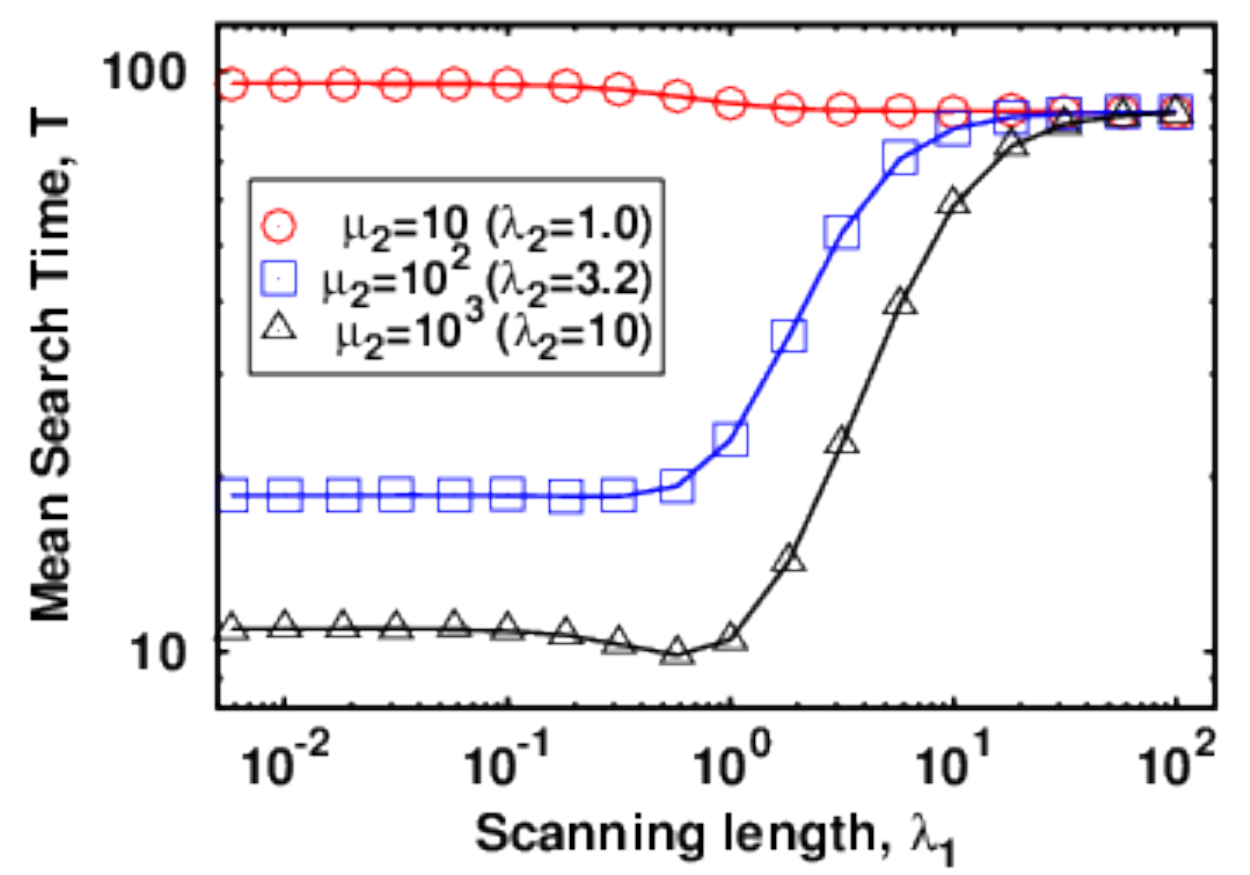}[A]
\includegraphics[width=7.5cm]{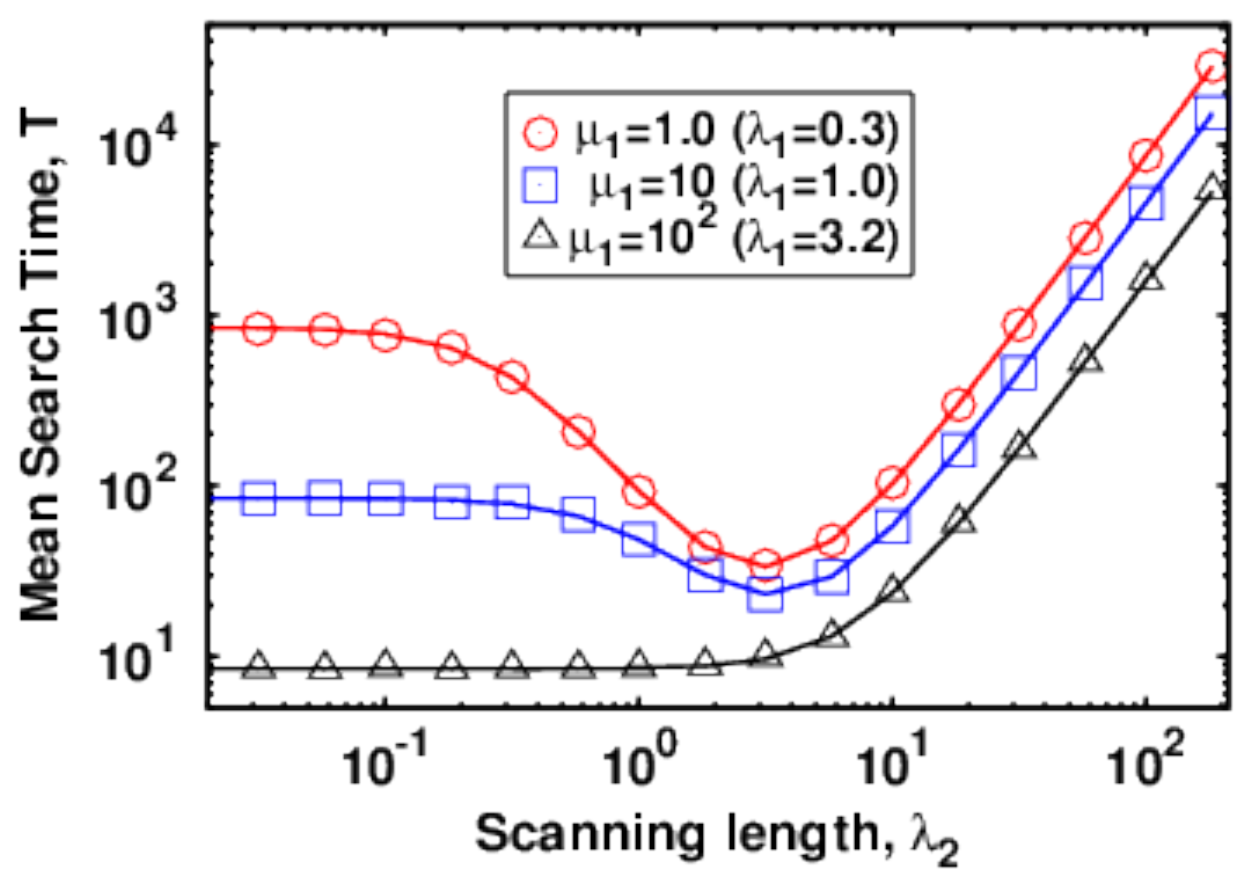}[B]
\caption{The position-averaged mean search time $T$ as a function of the scanning lengths $\lambda_1$ and $\lambda_2$. (A) Here $k_{\text{on}}=\mu_1=10$ and three different values of $\mu_2$ are utilized. We varied $k_{\text{off}}$ to change $\lambda_1$. (B) Here $\mu_2=100$, $k_{\text{off}}=10$ and three different values of $\mu_1$ are utilized. We varied $k_{\text{on}}$ to change $\lambda_2$. The solid lines are from the exact formulas and the symbols are from the Monte Carlo computer simulations. The lattice size is $2L+1=101$ and the target is at the center of the lattice.}
\label{fig-2}
\end{figure}

The results of our calculations for different ranges of kinetic parameters are presented in Figs. 2 and 3. We start with investigating the dependence of the search dynamics on the scanning lengths, and first we consider how the search time $T$ varies as a function of $\lambda_1$ in Fig. \ref{fig-2}A. One can identify three different dynamic regimes here. For $\lambda_1 > L$, the results converge and the mean search times become independent of the hopping rate in the search conformation $\mu_{2}$ and the switching rates between the conformations. This  result is easy to understand because for large $\lambda_{1}$ the searching molecule remains mostly in the recognition mode (state 1), and it can find the target without exploring the search conformation (state 2). Another dynamic behavior is observed for $\lambda_1 < 1$. Here the molecule spends most of the time in the search conformation (state 2), with very rare switchings to the recognition mode (state 1). This explains the independence of the mean search time from the scanning length $\lambda_{1}$. Also, in this regime the target can be reached mostly by the conformational transition from the site $n=0$ in the search mode, and the faster the hopping rate $\mu_{2}$ the higher the probability for this to happen because this site will be visited more frequently. The intermediate dynamic behavior is observed between these two limiting cases, when $1 < \lambda_{1} < L$, and the search involves scanning of the lattice in both conformations and frequent changes between them.

Analyzing the dependence of the mean search time on the scanning length $\lambda_{2}$, as shown in Fig 2B, three dynamic phases are again observed. For $\lambda_{2}<1$, no dependence on the scanning length is found because the searcher is mostly in the state 1 (recognition), and it can find the target without switching to another conformation. In addition, the faster you move in this mode (larger $\mu_{1}$), the sooner the target will be located. The dynamics is different for $\lambda_{2} >L$ because the system spends most of the time in the search conformation (state 2) from which the target cannot be located. Decreasing the switching rate $k_{\text{on}}$ (larger $\lambda_{2}$) will make finding the target even more difficult. But at the same time increasing the hopping rate $\mu_{1}$ will accelerate the search because after the rare switch to the recognition mode the target can be found before switching back to the search mode. The dynamic regime when $1 < \lambda_{2} < L$ exhibits the intermediate dynamic behavior between these two limiting cases, as expected. In this regime, the mean search time $T$ can be significantly shorter than those of two other regimes, and we argue that is because the searcher molecule can explore the space fast without much losing its recognition ability of the target. One can also see this from the fact that the most optimal conditions for the search are achieved for the conformational transitions rates that are not slow and too fast.

It is interesting to note that the conformational transitions dominate in the intermediate dynamic regimes ($1 < \lambda_{1}$,$\lambda_{2}<L$), but this does not always lead to the most optimal search dynamics: see the two upper curves in Fig. 2A and the lower curve in Fig. 2B. It can only happen if switching helps to explore the phase space more rapidly. This suggests that in the chemical and biological systems with the alternating dynamics there is an optimal range of the transition rates that can accelerate the search dynamics. Thus, in contrast to some widely expressed views, it is generally not valid to claim that just engaging in the stochastic switching will always accelerate the search dynamics.

To explore more the dynamic properties of the system with stochastic switchings between conformations, we calculate the dependence of the search times on the hopping rates $\mu_{1}$ and $\mu_{2}$, and the results are presented in Fig. 3. One can see that for relatively small values of $\mu_{1}$  (hopping rate in the recognition mode) the mean search times do not depend on them because in this case the system is mostly found in the search mode and the target can be reached via the stochastic transition at $n=0$ (Fig. 3A). Only for larger values of $\mu_1$ the target can be also reached from sliding along the lattice in the recognition mode, and increasing $\mu_{1}$ lowers the search time. A different behavior is observed for the dependence of $T$ as a function of $\mu_{2}$: see Fig. 3B. Increasing the hopping rate $\mu_{2}$ (the hopping rate in the search mode) accelerates the search because the system can explore faster the overall space, while for large $\mu_{2}$ the search dynamics becomes independent of the hopping rate. In this case, other processes (stochastic transition from the site $n=0$ to another conformation) is the rate-limiting step in the search process. It is also clear that  $T$ is independent of $\mu_{2}$ for large values of $\lambda_{1}$: see the lower curve in Fig. 3B. This corresponds to the situation when the molecule is almost always in the recognition mode and the target can be found without going into another conformational state.

Our theoretical analysis fully agrees with Monte Carlo computer simulations, and it suggests that the dynamic behavior in the system is determined by the relative values of the several length scales such as the scanning length $\lambda_{1}$, $\lambda_{2}$, the size of the system $L$, and the size of the target site 
(taken to be equal to unity). Combining these considerations together, we develop a general dynamic phase diagram for the molecular search with stochastic conformational transitions. The results are presented in Fig. 4. One can see that up to nine dynamic phases can be identified, showing a very rich behavior even in the relatively simple system with conformational transitions between only two modes.

\begin{figure}
\includegraphics[width=7.5cm]{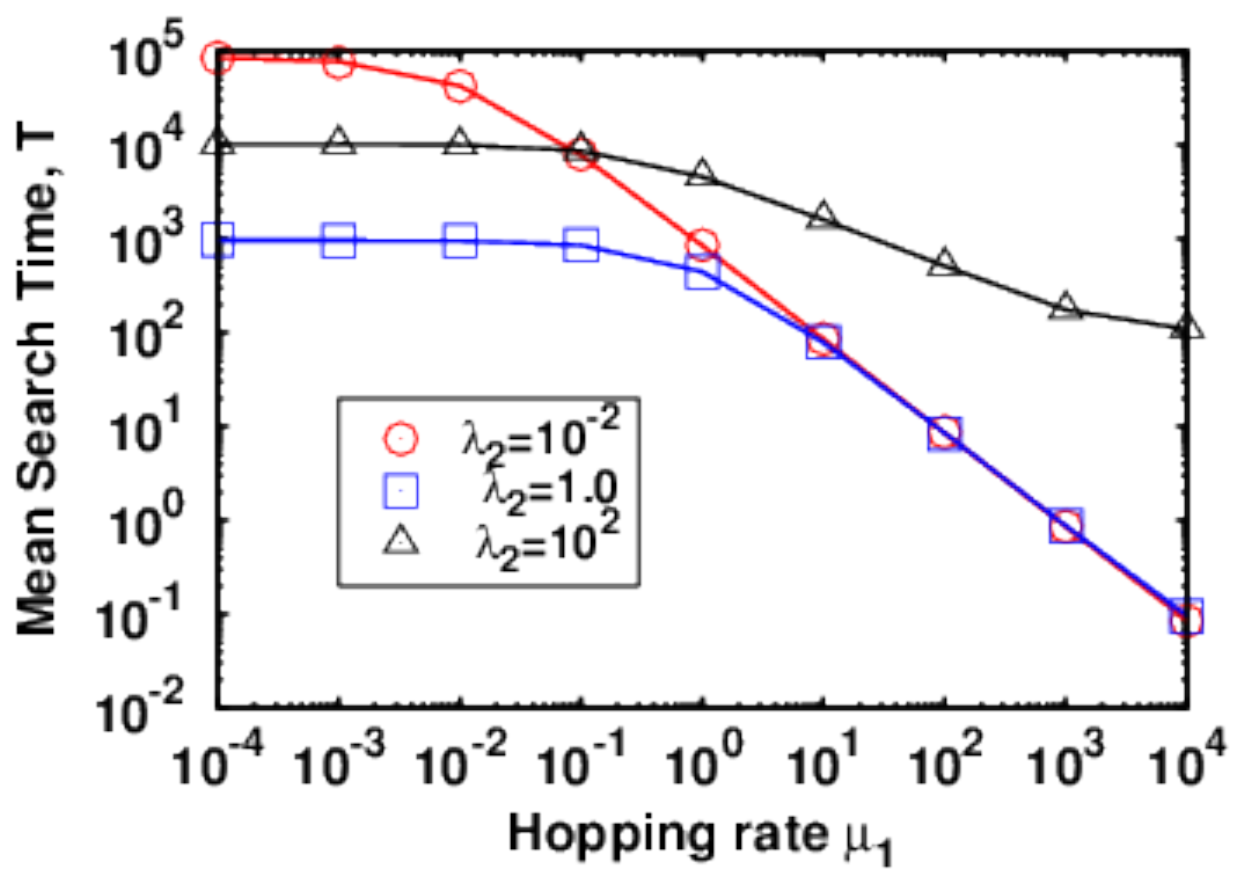}[A]
\includegraphics[width=7.5cm]{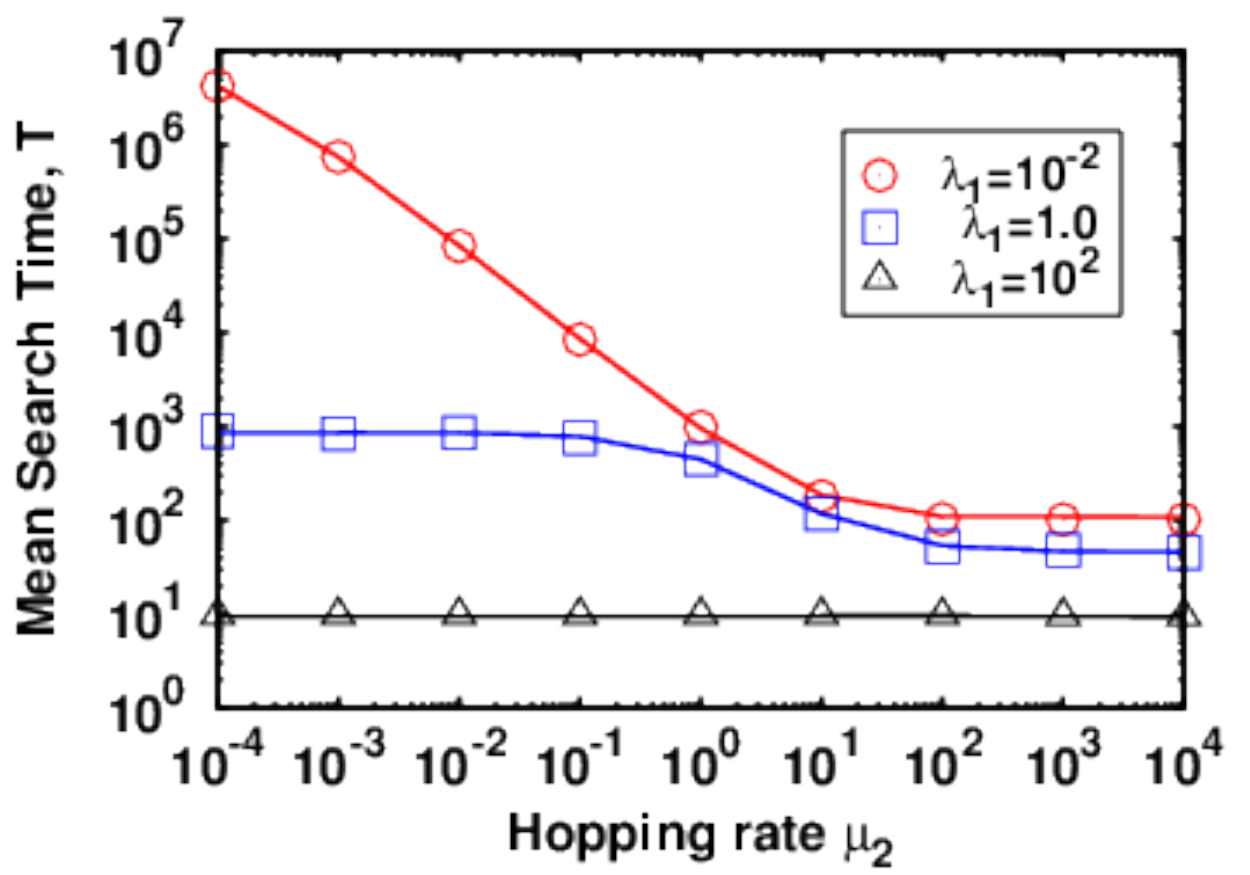}[B]
\caption{The position-averaged mean search times $T$ as a function of the hopping rates in different conformations. (A) Here $\mu_2=100$ and $k_{\text{off}}=1$ and three values of $k_{\text{on}}$ are utilized. (B) Here $\mu_1=100$ and $k_{\text{on}}=1$ and three values of $k_{\text{off}}$ are utilized. The solid lines are from the exact formulas and the symbols are from the Monte Carlo computer simulations. The lattice size is $2L+1=101$ and the target is at the center of the lattice.}
\label{fig-4}
\end{figure}


\section{Limiting Cases}
\label{approx}

Although our theoretical method provides explicit solutions for all ranges of kinetic parameters, to explain better the molecular search with alternating dynamics it is convenient to consider some limiting cases. In these situations, more transparent analysis can be done, which might clarify better the mechanisms of the search processes with stochastic transitions. There are two cases, corresponding to $\lambda_1 \gg \lambda_2$ and $\lambda_2 \gg \lambda_1$, which can be treated this way.

\begin{figure}
\includegraphics[width=7.5cm]{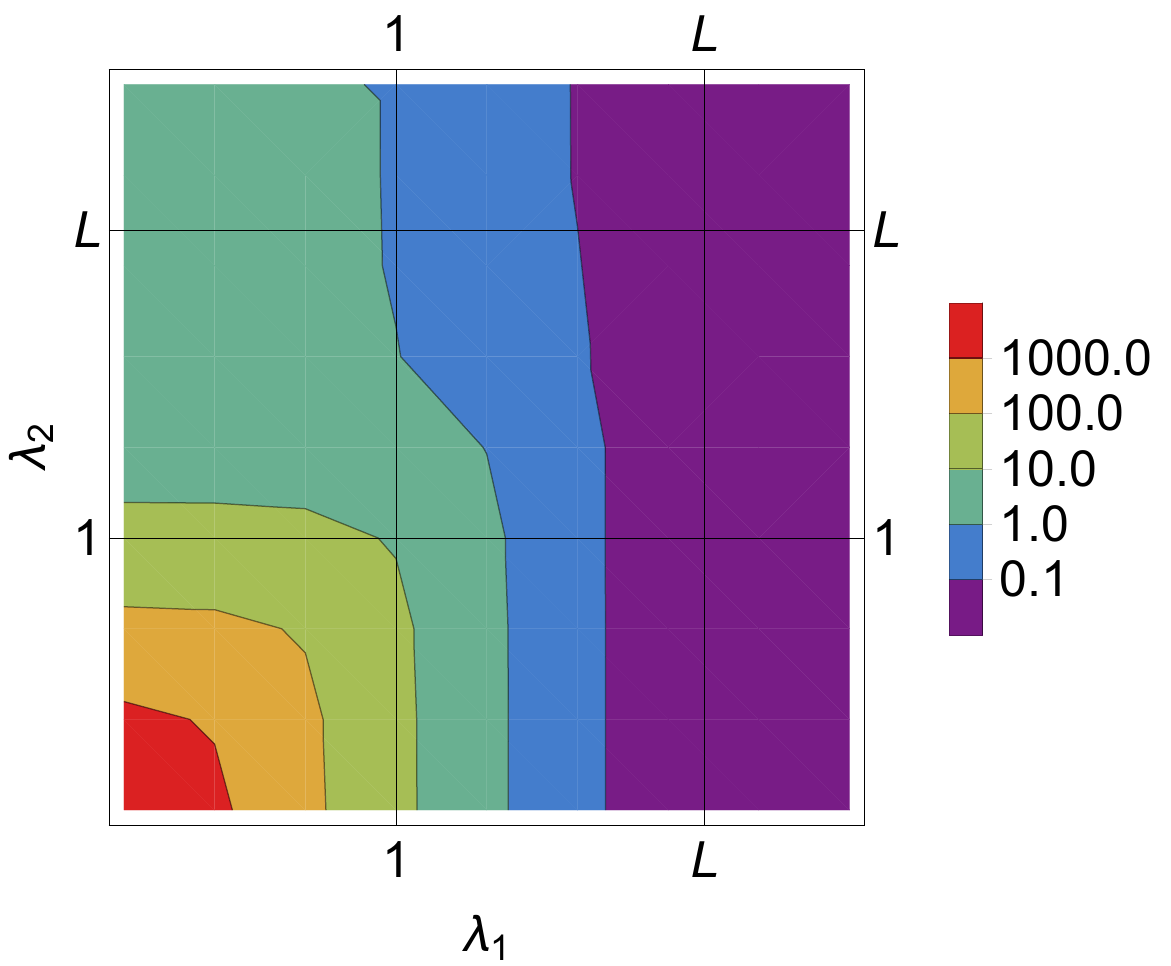}
\caption{Dynamic phase diagram for the molecular search with stochastic conformational transitions. The calculations used the lattice size $2L+1=101$, $k_{\text{on}}=k_{\text{off}}=100$, and we varied $\mu_1$ and $\mu_2$. The color map shows the values of the position averaged mean search times $T$. Lines schematically separate different dynamic regimes.}
\label{fig-3}
\end{figure}

\subsection{$\lambda_1 \gg \lambda_2$ case}

In this regime, the molecule spends most of the time in the recognition mode with occasional rare transitions to the search mode, in which it does not slide and switches back. Then the dynamics can be well described by assuming $\mu_{2}=0$. Using this result in Eq. (2) leads to \begin{equation}
	\frac{\partial F_2(n,t)}{\partial t}=k_{\text{on}}F_1(n,t)-k_{\text{on}}F_2(n,t),
\end{equation}
which in the Laplace domain can be written as
\begin{equation}
(s+k_{\text{on}})\widetilde{F_2(n,s)}=k_{\text{on}}\widetilde{F_1(n,s)}.
\label{eq4}
\end{equation}
Now analyzing Eqs. (3) and (4) for $\mu_{2}=0$ corresponds to solving a second-order difference equation instead of the much more complex fourth-order difference equation for the general situation. Similar analysis has been done in previous studies  on the protein search for targets on DNA \cite{veksler13,kochugaeva16,kochugaeva17}.

We assume that the general solution is of the form $\widetilde{F_1(n)} \simeq A x^n$, which leads to a quadratic equation, 
\begin{widetext}
\begin{equation}
\mu_1(s+k_{\text{on}})x^2-\left[s^{2}+s(2\mu_{1}+k_{\text{on}}+k_{\text{off}})+2\mu_{1}k_{\text{on}}\right]x+\mu_1(s+k_{\text{on}})=0,
\end{equation}
\end{widetext}
with roots given by
\begin{widetext}
\begin{equation}
x_{1}=\frac{s^{2}+s(2\mu_{1}+k_{\text{on}}+k_{\text{off}})+2\mu_{1}k_{\text{on}}-\sqrt{D}}{2\mu_{1}(s+k_{\text{on}})}, \quad x_{2}=1/x_{1};
\end{equation}
\end{widetext}
while the parameter $D$ is equal to
\begin{widetext}
\begin{equation}
D=\left[s^{2}+s(2\mu_{1}+k_{\text{on}}+k_{\text{off}})+2\mu_{1}k_{\text{on}}\right]^{2}-4\mu_{1}^{2}(s+k_{\text{on}})^{2}. 
\end{equation}
\end{widetext}
Then the general solution is $\widetilde{F_1(n)}=A_1 x_1^n+A_2 x_1^{-n}$, where $x_1$ is given by  Eq. (10), and $A_1$ and $A_2$ are the unknown coefficients that can be determined from the boundary conditions. The final expression for the first-passage probability function (in the Laplace form) is
\begin{equation}
\widetilde{F_1(n)}=\frac{x_1^{2L+n}}{x_1^{2L}+1}+\frac{x_1^{-n}}{x_1^{2L}+1}.
\end{equation}
This yields the following expression for the mean search time from the site $n$ in the recognition conformation, 
\begin{equation}
T_1(n)=\frac{\left|n\right|(2L+1-\left|n\right|)}{2\mu_{\text{eff1}}},
\end{equation}
where $\mu_{\text{eff1}}=\mu_1/(1+k_{\text{off}}/k_{\text{on}})$. This result has a simple physical interpretation. It describes the mean first-passage time of finding the target (which is at the origin) starting from the site $n$ by purely 1D motion in the segment of size $(2L+1)$ with the effective  hopping rate $\mu_{\text{eff1}}$.  In this regime, the molecule explores the lattice mostly in the search conformation with the occasional switchings to the immobile recognition mode. At large times, the system reaches the effective equilibrium between two conformations because the particle does not slide in the recognition mode. Therefore the hopping rate is rescaled by an equilibrium fraction of finding the molecule in the search configuration, $f=\frac{k_{\text{on}}}{k_{\text{on}}+k_{\text{off}}}$. 

Finally, it can be shown the position-average mean search time is
\begin{equation}
T=\frac{1}{k_{\text{on}}}+\frac{1}{2L+1}\frac{1}{\mu_{\text{eff1}}}\sum_{n=1}^{L}n(2L+1-n).
\label{eq-20}
\end{equation}
The first term corresponds to the transition time from the state 2 to the state 1, and the second term corresponds to 1D search time with effective hopping rate $\mu_{\text{eff1}}$. It can be shown that these results fully agree with exact calculations in the limit of very small $\lambda_{2}$.

\subsection{$\lambda_1 \ll \lambda_2$ case}

Exact calculations can be also performed in the opposite limit when the searching molecule moves along the lattice mostly in the search mode with occasional switchings to the recognition mode. In this regime, the target is reached only via the conformational transitions from the site $n=0$, and we can approximate the dynamics as $\mu_1=0$, modifying Eq. (1) as
\begin{equation}
\frac{\partial F_1(n,t)}{\partial t}=k_{\text{off}}F_2(n,t)-k_{\text{off}}F_1(n,t).
\end{equation}
Again using the Laplace transformations, the mean search time $T$ can be explicitly evaluated. The final expression is
\begin{equation}
T_2(n)=\frac{1}{2\mu_{\text{eff2}}} \left|n\right|(2L-\left|n\right|+1)+\left[\frac{2L+1}{k_{\text{on}}}+\frac{2L}{k_{\text{off}}}\right],
\end{equation}
where $\mu_{\text{eff2}}=\mu_2 /(1+k_{\text{on}}/~ k_{\text{off}})$. Here the first term describes  the mean time to travel from the initial position $n$ to the site $n=0$ in the search conformation, and the second term corresponds to the time of conformational transition from the state 2 to the state 1 and return back to the search conformation. The number of transitions from the state 2 to the state 1 is $2L+1$, while the number of reversed transitions is less by one because the last transition to the recognition mode is going to be directly to the target, which ends the search process. It can be shown explicitly that these results agree with the full solutions in this limit.

\section{Summary and Conclusions}
\label{conclusion}

We presented a comprehensive theoretical analysis of the one-dimensional discrete-state stochastic model of the molecular search with random conformational transitions. Stimulated by biological processes of protein-DNA interactions, this model also serves as a general testing ground for understanding the role of the intermittent search phenomena in complex natural systems. Using the method of first-passage probabilities, we were able to obtain the explicit quantitative description of the dynamics for all ranges of kinetic parameters. It has been argued that that four length scales specify the dynamic phase diagram in this system. They include the size of the target site, the size of the system, and the two scanning lengths in different conformations. A general dynamic phase diagram  that describes all possible search behaviors is constructed. It is found that the stochastic transitions between two conformations might optimize the search dynamics at the conditions when the switching leads to a more rapid exploration of the phase space without too much losing the recognition ability. Our theoretical calculations also show that the acceleration does not always happen, indicating that the intermittent search is not always the most efficient dynamic regime. In addition, because our model is one-dimensional, any changes in the dynamic properties cannot be associated with the ``lowering of dimensionality" arguments that are widely expressed in the literature for the intermittent search phenomena \cite{adam68}.

Although the presented theoretical model gives a clear picture of how the stochastic switching is affecting the dynamics, it is important to note that its application to real biological systems is rather limited. One needs to include the coupling to the motion in other dimensions (3D in the bulk and/or 2D on the surface) to make the description more realistic \cite{holcman11,holcman15,kochugaeva16}. At the same time, the presented theoretical framework clarifies many aspects of the mechanisms of the intermittent search. It will be interesting to test these theoretical predictions in experimental systems as well as in the more advanced theoretical descriptions.

\section*{Acknowledgements}
The work was supported by the Welch Foundation (C-1559), by the NSF (CHE-1664218), and by the Center for Theoretical Biological Physics sponsored by the NSF (PHY-1427654).

\section*{Appendix: Details of Calculations}

Here we describe the details of the first-passage probabilities calculations to evaluate the search dynamics. We define the first-passage time probability density to find the target with the initial position $n$ at time $t=0$ in state $i$ ($=1$ or $2$) as $F_i(n, t)$. The number of lattice sites is $2L+1$, labeled as $-L, -L+1, ..., 0, ..., L-1, L$. We consider the target at the center of the lattice, $n=0$. The evolution of $F_i(n,t)$ follows the following backward master equations \cite{veksler13,kochugaeva16}
\begin{widetext}
\begin{equation}
	\frac{\partial F_1(n,t)}{\partial t}=\mu_1\left[F_1(n-1,t)+F_1(n+1,t)\right]+k_{\text{off}}F_2(n,t)-(2\mu_1+k_{\text{off}})F_1(n,t),
\end{equation}
\end{widetext}
\begin{widetext}
\begin{equation}
	\frac{\partial F_2(n,t)}{\partial t}=\mu_2[F_2(n-1,t)+F_2(n+1,t)]+k_{\text{on}}F_1(n,t)-(2\mu_2+k_{\text{on}})F_2(n,t),
\end{equation}
\end{widetext}
for $-L<n< +L$. At the boundary $n=-L$ we have
\begin{widetext}
\begin{equation}
	\frac{\partial F_1(-L,t)}{\partial t}=\mu_1F_1(-L+1,t)+k_{\text{off}}F_2(-L,t)-(\mu_1+k_{\text{off}})F_1(-L,t),
\end{equation}
\end{widetext}
\begin{widetext}
\begin{equation}
	\frac{\partial F_2(-L,t)}{\partial t}=\mu_2F_2(-L+1,t)+k_{\text{on}}F_1(-L,y)-(\mu_2 +k_{\text{on}})F_2(-L,t).
\end{equation}
\end{widetext}
Similar equations can be written for $n=L$.

Now the Laplace transform, $\widetilde{F_i(n,s)}=\int_{0}^{\infty}F_i(n,t)\exp(-st)dt$, where $s$ is the Laplace variable, can be applied for all master equations to transform them from the differential equations into algebraic equations. This leads to
\begin{equation}
(s+2\mu_1+k_{\text{off}})\widetilde{F_1(n)}=\mu_1[\widetilde{F_1(n-1)}+\widetilde{F_1(n+1)}]+k_{\text{off}}\widetilde{F_2(n)},
\label{eq3}
\end{equation}
\begin{equation}
(s+2\mu_2+k_{\text{on}})\widetilde{F_2(n)}=\mu_2[\widetilde{F_2(n-1)}+\widetilde{F_2(n+1)}]+k_{\text{on}}\widetilde{F_1(n)}.
\label{eq4}
\end{equation}
These expressions form a system of two coupled second-order difference equations. To solve this system, these two equations  can be combined to make them a single fourth-order difference equation. This procedure yeilds,
\begin{equation}
\widetilde{F_1(n)}=\frac{(s+2\mu_2+k_{\text{on}})}{k_{\text{on}}}\widetilde{F_2(n)}-\frac{\mu_2}{k_{\text{on}}}[\widetilde{F_2(n-1)}+\widetilde{F_2(n+1)}].
\end{equation}
Substituting this into Eq.\ref{eq3} yields,
\begin{widetext}
\begin{equation}\label{eq24}
	(a_1a_2+2\mu_1\mu_2-k_{\text{on}}k_{\text{off}})\widetilde{F_2(n)}=(a_1\mu_2+a_2\mu_1)[\widetilde{F_2(n-1)}+\widetilde{F_2(n+1)}]-\mu_1\mu_2[\widetilde{F_2(n-2)}+\widetilde{F_2(n+2)}],
\end{equation}
\end{widetext}
where new auxiliary functions are defined as $a_1\equiv s+2\mu_1+k_{\text{off}}$ and $a_2\equiv s+2\mu_2+k_{\text{on}}$. Now, introducing $A\equiv a_1a_2+2\mu_1\mu_2-k_{\text{on}}k_{\text{off}}$, $B \equiv a_1\mu_2+a_2\mu_1$, and $C \equiv \mu_1\mu_2$, we can rewrite Eq. (\ref{eq24}) in the more compact form
\begin{equation}
	A \widetilde{F_2(n)}=B [\widetilde{F_2(n-1)}+\widetilde{F_2(n+1)}]-C[\widetilde{F_2(n-2)}+\widetilde{F_2(n+2)}].
	\label{eq7}
\end{equation}
To solve this equation, we assume that the solution is of the form $\widetilde{F_2(n)}=\alpha x^n$, and substituting this into the Eq. (\ref{eq7}) gives a quartic equation,
	\begin{equation}
	C x^4-B x^3+A x^2-B x+C=0.
	\end{equation}
It can be shown that it has four  roots of the form $(x_1$, $1/x_1$, $x_2$, $1/x_2)$, and $x_{i}$ ($i=1$ or $2$) might be a complex number. Then the general solution is $\widetilde{F_2(n)}=A_1 x_1^n+A_2 x_1^{-n}+B_1 x_2^n+B_2 x_2^{-n}$. We need to determine the four unknown coefficients ($A_{1}$, $A_{2}$, $B_{1}$ and $B_{2}$) by using the following boundary conditions:

(1) At $n=-L$ the corresponding master equations in Laplace domain are
\begin{equation}
[s+\mu_1+k_{\text{off}}]\widetilde{F_1(-L)}=\mu_1 \widetilde{F_1(-L+1)}+k_{\text{off}}\widetilde{F_2(-L)};
\label{eq9}
\end{equation}
\begin{equation}
[s+\mu_2+k_{\text{on}}]\widetilde{F_2(-L)}=\mu_2 \widetilde{F_2(-L+1)}+k_{\text{on}}\widetilde{F_1(-L)}.
\label{eq10}
\end{equation}
Substituting Eq.\ref{eq10} into Eq. \ref{eq9} gives
\begin{widetext}
\begin{equation}
\begin{split}
	[s+\mu_1+k_{\text{off}}]\big[(s+\mu_2+k_{\text{on}})\widetilde{F_2(-L)}-\mu_2 \widetilde{F_2(-L+1)} \big]= \\
	\mu_1 \big[ (s+2\mu_2+k_{\text{on}})\widetilde{F_2(-L+1)}-\mu_1(\widetilde{F_2(-L)}+\widetilde{F_2(-L+2)} ) \big]+k_{\text{on}}k_{\text{off}}\widetilde{F_2(-L)}
\end{split}
\end{equation}
\end{widetext}
This can be written as
\begin{equation}
\begin{split}
\big[ (s+\mu_1+k_{\text{off}})(s+\mu_2+k_{\text{on}})+\mu_1\mu_2-k_{\text{on}}k_{\text{off}}  \big] \widetilde{F_2(-L)}= \\
\big[ (s+\mu_1+k_{\text{off}})\mu_2+\mu_1 a_2)  \big] \widetilde{F_2(-L+1)}- \mu_1\mu_2 \widetilde{F_2(-L+2)}
\end{split}
\label{bc1}
\end{equation}
This is the first boundary condition.

(2)The master equations for the site $n=-L+1$ in two conformational states are
\begin{equation}
a_1 \widetilde{F_1(-L+1)}=\mu_1 [ \widetilde{F_1(-L)}+\widetilde{F_1(-L+2)}]+k_{\text{off}}\widetilde{F_2(-L+1)};
\label{eq13}
\end{equation}
\begin{equation}
a_2 \widetilde{F_2(-L+1)}=\mu_2 [\widetilde{F_2(-L)}+\widetilde{F_2(-L+2)}]+k_{\text{on}}\widetilde{F_1(-L+1)}.
\label{eq14}
\end{equation}
We can also rewrite the master equations for the sites $n=-L$, $n=-L+1$ and $n=-L+2$,
	\begin{equation}
	\begin{split}
	k_{\text{on}}\widetilde{F_1(-L+1)}=a_2 \widetilde{F_2(-L+1)}-\mu_2[\widetilde{F_2(-L)}+\widetilde{F_2(-L+2)}], \\
	k_{\text{on}}\widetilde{F_1(-L)}=(s+\mu_2+k_{\text{on}})\widetilde{F_2(-L)}-\mu_2 \widetilde{F_2(-L+1)}, \\
	k_{\text{on}}\widetilde{F_1(-L+2)}=a_2 \widetilde{F_2(-L+2)}-\mu_2[\widetilde{F_2(-L+1)}+\widetilde{F_2(-L+3)}]. \\
	\end{split}
	\end{equation}
Substituting them into Eq. \ref{eq13} yields
\begin{widetext}
\begin{equation}
\begin{split}
a_1\big[ a_2 \widetilde{F_2(-L+1)}-\mu_2 \{ \widetilde{F_2(-L)}+\widetilde{F_2(-L+2)} \} \big]= \\
\mu_1 \big[ (s+\mu_2+k_{\text{on}}) \widetilde{F_2(-L)}-\mu_2 \widetilde{F_2(-L+1)} + a_2 \widetilde{F_2(-L+2)}- \mu_2 \{ \widetilde{F_2(-L+1)}+\widetilde{F_2(-L+3)} \} \big]+k_{\text{on}}k_{\text{off}}\widetilde{F_2(-L+1)},
\end{split}
\end{equation}
\end{widetext}
which can be simplified as
\begin{widetext}
\begin{equation}
\begin{split}
\big[a_1\mu_2+(s+\mu_2+k_{\text{on}})\mu_1 \big]\widetilde{F_2(-L)}-\big[a_1 a_2+2 \mu_1\mu_2-k_{\text{on}}k_{\text{off}} \big]\widetilde{F_2(-L+1)}+\big[a_1 \mu_2+a_2\mu_1\big]\widetilde{F_2(-L+2)}-\mu_1\mu_2 \widetilde{F_2(-L+3)}=0.
\end{split}
\end{equation}
\end{widetext}
This is the second boundary condition.
	
(3) The additional boundary condition is associated with the dynamics at the target site because $\widetilde{F_1(n=0)}=1$. The Eq. \ref{eq4} with $n=0$ is
	\begin{equation}
	a_2 \widetilde{F_2(0)}=\mu_2(\widetilde{F_2(-1)}+\widetilde{F_2(1)})+k_{\text{on}}.
	\end{equation}
Because of the symmetry of the system, $\widetilde{F_2(-1)}=\widetilde{F_2(1)}$, so that the above equation simplifies into
	\begin{equation}
	a_2 \widetilde{F_2(0)}=2 \mu_2 \widetilde{F_2(-1)}+k_{\text{on}}.
	\end{equation}
This is the third boundary condition.

(4) From master equations for $n=-1$ and $n=-2$ one can derive
\begin{equation}
\begin{split}
k_{\text{on}}\widetilde{F_1(-1)}=a_2\widetilde{F_2(-1)}-\mu_2\{ \widetilde{F_2(-2)}+\widetilde{F_2(0)} \}, \\
k_{\text{on}}\widetilde{F_1(-2)}=a_2\widetilde{F_2(-2)}-\mu_2 \{ \widetilde{F_2(-3)}+\widetilde{F_2(-1)} \}.
\end{split}
\label{eq20}
\end{equation}
From Eq. (\ref{eq3}) with $n=-1$ we have
\begin{equation}
a_1 \widetilde{F_1(-1)}=\mu_1 \{\widetilde{F_1(-2)}+\widetilde{F_1(0)}\}+k_{\text{off}}\widetilde{F_2(-1)}.
\label{eq21}
\end{equation}
Substituting Eq. (\ref{eq20}) into Eq. (\ref{eq21}) yields
\begin{widetext}
\begin{equation}
a_1 \big[ a_2\widetilde{F_2(-1)}-\mu_2\{ \widetilde{F_2(-2)}+\widetilde{F_2(0)}\} \big]=\mu_1\big[ a_2\widetilde{F_2(-2)}-\mu_2\{ \widetilde{F_2(-3)}+\widetilde{F_2(-1)}\} \big] +\mu_1 k_{\text{on}}+k_{\text{on}}k_{\text{off}}\widetilde{F_2(-1)},
\end{equation}
\end{widetext}
which can be further simplified as
\begin{widetext}
\begin{equation}
\mu_1\mu_2\widetilde{F_2(-3)}-(a_1 \mu_2+a_2\mu_1)\widetilde{F_2(-2)}+(a_1 a_2+\mu_1 \mu_2-k_{\text{on}}k_{\text{off}})\widetilde{F_2(-1)}-a_1\mu_2\widetilde{F_2(0)}=\mu_1 k_{\text{on}}.
\end{equation}
\end{widetext}
This is the fourth boundary condition.

These boundary conditions are considered together, and numerical  solutions are obtained for the unknown coefficients $A_{1}$, $A_{2}$, $B_{1}$ and $B_{2}$. Then Eq. (5) is utilized for explicit calculations of  the mean search times.

\end{document}